\documentclass[letterpaper, 10 pt, conference]{ieeeconf}  
\IEEEoverridecommandlockouts                             
\newtheorem{corollary}{Corollary}

\newtheorem{proposition}{Proposition}

\newtheorem{assumption}{Assumption}
\usepackage{amssymb}
\usepackage{epsfig}
\usepackage{psfrag}
\usepackage{algorithm}
\usepackage{algpseudocode}
\usepackage{array}
\usepackage{epstopdf}
\usepackage{tikz, subcaption, cite}
\usepackage{mathtools}
\usepackage{slashbox}
\usepackage{comment}
\usepackage{tabularx}
\usepackage{adjustbox}

\renewcommand{\theenumi}{\roman{enumi}}%

\usepackage[none]{hyphenat}

\title{\LARGE \bf
A Quantal Response Analysis of Defender-Attacker Sequential Security Games}
\author{Md Reya Shad Azim, and~Mustafa Abdallah% <-this % stops a space
\thanks{This research was supported by EMPOWER grant from the Vice Chancellor for Research at Indiana University-Purdue University Indianapolis. Md Reya Shad Azim is with the Department of Electrical and Computer Engineering at Indiana University-Purdue University Indianapolis. Email: {\tt azimm@iu.edu}. Mustafa Abdallah is with the Department of Computer and Information Technology at Indiana University-Purdue University Indianapolis. Email: {\tt{abdalla0@iu.edu}}.}.%
}

\begin{document}

\sloppy

\maketitle
\thispagestyle{empty}
\pagestyle{empty}

%%%%%%%%%%%%%%%%%%%%%%%%%%%%%%%%%%%%%%%%%%%%%%%%%%%%%%%%%%%%%%%%%%%%%%%%%%%%%%%%
\begin{abstract}
We explore a scenario involving two sites and a sequential game between a defender and an attacker, where the defender is responsible for securing the sites while the attacker aims to attack them. Each site holds a loss value for the defender when compromised, along with a probability of successful attack. The defender can reduce these probabilities through security investments at each site. The attacker's objective is to target the site that maximizes the expected loss for the defender, taking into account the defender's security investments. While previous studies have examined security investments in such scenarios, our work investigates the impact of bounded rationality exhibited by the defender, as identified in behavioral economics. Specifically, we consider quantal behavioral bias, where humans make errors in selecting efficient (pure) strategies. We demonstrate the existence of a quantal response equilibrium in our sequential game and analyze how this bias affects the defender's choice of optimal security investments. Additionally, we quantify the inefficiency of equilibrium investments under quantal decision-making compared to an optimal solution devoid of behavioral biases. We provide numerical simulations to validate our main findings.
\end{abstract}

%%%%%%%%%%%%%%%%%%%%%%%%%%%%%%%%%%%%%%%%%%%%%%%%%%%%%%%%%%%%%%%%%%%%%%%%%%%%%%%%
\section{Introduction}

Enhancing the security of cyber-physical systems (CPS) against sophisticated adversaries presents a formidable challenge~\cite{humayed2017cyber}. In such contexts, adversaries often exploit vulnerabilities to target specific objectives, while defenders typically contend with limited resources for vulnerability mitigation~\cite{laszka2015survey,alpcan2010network}. Game-theoretic models have been leveraged to capture these settings under various assumptions regarding the strategies available to defenders and attackers~\cite{farrow2007economics,guan2017modeling,powell2007allocating}. Of particular relevance to our study is the work by Powell et al.~\cite{powell2007allocating}, which examined a sequential framework involving defenders and attackers, elucidating optimal strategies for each player.

A common thread in much of the existing work for securing CPS is that the defenders and attackers are assumed to behave according to classical models of fully rational decision-making, taking actions to minimize their expected loss (or maximizing expected utility). However, a large body of work in behavioral economics and psychology has shown that humans consistently deviate from such classical models of decision-making.  For example, {\it quantal response} research showed that humans consistently make errors in choosing efficient (pure) strategies when making decisions~\cite{dhami2016foundations}.

Notably, the quantal response equilibrium takes into account the fact that human decision-makers may not always make perfectly rational decisions. Instead, it allows for decision-making where the players in a game may not always choose the best response with certainty, but their choices are probabilistic and influenced by factors such as noise and cognitive limitations. Many empirical studies have provided evidence for this class of behavioral models~\cite{anderson2004noisy,zhuang2014stability,morgan1992analysis}.

While a substantial body of literature on quantal response exists in economics and psychology, the existing research exploring the impact of such behavioral decision-making on CPS security and robustness primarily draws from psychological studies \cite{anderson2007information} and human subject experiments \cite{redmiles2018dancing,baddeley2011information}. However, these studies often lack rigorous mathematical models of players' behavior. Recently, there has been a growing trend towards leveraging mathematical analysis to model and predict the effect of behavioral decision-making on players' investments \cite{8814307, 9030279,sanjab2017prospect,abdallah2020effect}. However, these works have only focused only on prospect-theoretic attitudes and have not considered quantal behavioral errors which is the focus of our work.

%is said to be a QRE when players assign probabilities to their available security investment profiles, 

In this paper, we introduce quantal response into a game-theoretic framework involving one defender and one attacker. We consider the case where the defender places her investments to best protect her sites, accounting for a strategic attacker who chooses which site to compromise to maximize the expected loss of the defender. We first show that such a game with behavioral players (with quantal responses) has a quantal response equilibrium (QRE). We then show that the probability of choosing the best investment strategy is non-decreasing in the behavioral level of the defender. We then characterize the main impacts of the behavioral level and the assets' losses on the security investments made by the defender. %, and on the decisions of the attacker. 
 
 We introduce a formal setting for quantal behavioral decision-making, described in further detail in the next two sections. We first show the uniqueness of the optimal defense allocation of the defender.  We then characterize the impacts of quantal response behavioral level on choosing the best investment decision by the defender; in particular, we show that the defender tends to choose her optimal investments with lower probability under higher level of quantal bias which increases her loss when attacked. We then show that the choice of the optimal defense strategy depends on the sites' losses and other available strategies. Finally, we introduce the notion of Price of Quantal Anarchy (PoQA) to quantify the inefficiency of the behavioral defender's choice of less efficient investments on her  expected loss and provide bounds on the PoQA. We finally provide numerical simulations to illustrate our findings.
%% Commented for CDC Final Version
%% We then show how a (non-behavioral) attacker's decision and payoff are affected by different behavioral levels for the defender.
\section{The Defender-Attacker Sequential Game}\label{sec:background}

In this section, we describe our sequential game framework, which builds upon the model introduced in  \cite{powell2007allocating}.

\subsection{Defender-Attacker Sequential Game Setup}

\textbf{Game Setup:} We consider a security game which consisting of two players, a defender and an attacker.  There are two {\it sites}, denoted site $1$ and site $2$, which the defender is trying to protect (and the attacker is trying to compromise). The defender has a budget $R \in \mathbb{R}_{>0}$ that she can spend on defending the both sites.  In particular, the defender moves first  and allocates an amount $r \in [0,R]$ to site $1$ and an amount $R-r$ to site $2$.  We assume that the attacker can observe the allocations made by the defender to each of the sites, after which she targets one of the two sites to attack.

\textbf{Defense Strategy Space:} The defense strategy space of the defender is defined by 
\vspace{-2mm}
\begin{equation*}
\vspace{-1mm}
X := \{\mathbf{r} \in \mathbb{R}^{2}_{\geq0} | 1^T \mathbf{r} = R  \, \}.    
\end{equation*}
In words, $X$ is the set of feasible security investments for the defender. It consists of all possible non-negative investments on the two sites such that the sum of these investments equals the defender's security budget $R$. This captures the real-world assumption of limited security resources for human security decision-makers. We denote any particular vector of investments by the defender as $\mathbf{r} \in X$, where $\mathbf{r} = [r, R-r]$.

\textbf{Probability of Successful Attack:} The probability of successful attack on site $1$, when the defense investment on that site is $r$, is denoted by $p_{1}(r)$.  We assume $p'_{1} < 0$ and $p''_{1} > 0$.  Similarly, for site 2, let $p_{2}(x)$ be the probability that an attack on site 2 succeeds if the defender spends $x=R-r$ defending that site. Once again, we assume that $p'_{2} < 0$ and $p''_{2} > 0$. In words, the probability that the attacker successfully compromises the site that it targets is a decreasing function of the amount invested in protecting that site by the defender. We emphasize that these assumptions are common in security games literature (e.g., \cite{powell2007allocating,9654432}). % baryshnikov2012security

\iffalse
where $x_i$ is the investment of defender at site $i$, the initial attack probability is $p_i^0$ and $s_i$ is the sensitivity of the site $i$. For the simplicity, we assume that the initial attack probability $p^0$ and the site sensitivity $s_i$ is $1$.
\fi

\textbf{Defender's Loss and Attacker's Gain:} The defender suffers a loss of one if site 1 is successfully attacked, and a loss of $A > 0$ if site 2 is successfully attacked. After the defender allocates her resources, the attacker targets the site that will maximize the defender's expected loss.  Thus, the defender's expected loss if it allocates $r$ to site 1 is given by
\begin{equation}\label{eq: Non_Behavioral_defender_expected_loss} 
L(\mathbf{r}) = \text{max} \left\{p_{1}(r), A p_{2}(R - r)\right\}.    
\end{equation}
This is also the attacker's expected gain. In other words, the attacker chooses which site to attack in order to maximize the defender's {\it true} expected loss. Thus, the attacker's action under a defense investment of $r$ on site 1 will yield the utility
\begin{equation}\label{eq: Behavioral_attacker_utility}
U(\textbf{r}) = \text{max} \{p_{1}(r), A p_{2}(R - r)\}.
\end{equation}

\begin{figure}
\begin{center}
\includegraphics[width=0.8\linewidth]{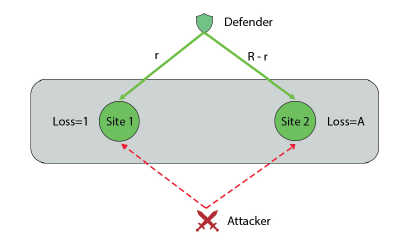}
\vspace{-2mm}
  \caption{A sequential game setup where the defender invests $r$ on site-1 and $R-r$ in site-2 (solid green arrows). The attacker attacks either site-1 or site-2 (dashed red arrows) after observing defender's investments on the two sites.}
  \label{fig:Sequential game setup}
\end{center}
\vspace{-6mm}
\end{figure}

Figure~\ref{fig:Sequential game setup} illustrates an illustrative example of our sequential game setup, which serves as the basis for evaluating our analysis throughout this paper. 

\subsection{Uniqueness of defender’s optimal investment
strategy}
\begin{proposition}\label{prop: equilibria_behavioral_complete_info}
%Subgame perfect equilibria of the Behavioral Defender-Attacker Sequential Game exist, and the defender's allocation is the same in all of them. This allocation, $\hat{r}(v)$,
%minmaxes the attacker;\footnote{SS: I don't know if we should say that this minimaxes the attacker, since the defender does not know the attacker's actual utility function.  The defender is just trying to minimize it's own loss function. Mus: Removed it.}
%is the unique solution to $\displaystyle \min_{r\in [0,R]} \max\{w(p_{1}(r, v)), A w(p_{2}(R - r))\}$.
The defender's optimal allocation to site $1$ to minimize $L(\mathbf{r})$ in \eqref{eq: Non_Behavioral_defender_expected_loss} is unique, and denoted by $r^*$.
\end{proposition}

\begin{proof}
First, note that by our assumption $p_1(r)$ is strictly decreasing in $r$.  Similarly, by our assumption that $p_2(x)$ is strictly decreasing in its argument, the function $p_2(R-r)$ is strictly increasing in $r$.  Thus, $A p_2(R-r)$ is strictly increasing in $r$.

First, consider the case where
$p_{1}(r) < A p_{2}(R-r) \forall r \in [0,R]$. Then, ${r}^* = 0$ is the unique solution that minimizes the defender's expected loss (since if the defender deviates to any investment $r > 0$, the defender's expected loss $A p_{2}(R - r)$ would be larger as $Ap_2(R-r)$ is increasing in $r$. %$\frac{dw(p_2(R-r))}{dr} > 0$). 

Second, consider the case in which  $p_{1}(r) > A p_{2}(R-r)) \forall r \in [0,R]$. Then, ${r}^* = R$ is the unique solution that minimizes the defender's expected loss (since if the defender deviates to any investment $r < R$, the defender's expected loss $p_{1}(r)$ would increase as $p_1(r)$ is decreasing in $r$ as shown earlier due to this second case). 

Now, consider the cases that are not considered by the above two (corner) cases. Note that since $p_1(r)$ is strictly decreasing in $r$, and since $Ap_2(R-r)$ is strictly increasing in $r$, and since neither function is always larger than the other (which was captured by the above two cases), there must be a unique value of $r$ at which $p_{1}(r) = A p_{2}(R - r)$.  This will be the unique solution to the problem of minimizing \eqref{eq: Non_Behavioral_defender_expected_loss}, which we denote by $r^*$.
\end{proof}

\begin{corollary}\label{cor:best case}
Suppose that the quantities $p_1(r)$ and $Ap_2(R-r)$ are such that $p_1(0) > Ap_2(R)$ and $p_1(R) < Ap_2(0)$. Let $p_1(r) = \exp(-r)$ and $p_{2}(r)=e^{-(R-r)}$, respectively. Then, the defender's optimal allocation to site $1$ in order to minimize $L(\mathbf{r})$ in \eqref{eq: Non_Behavioral_defender_expected_loss} is given by $r^* = \frac{R-\log A}{2}$.
\end{corollary}

\begin{proof}
Since  $p_1(r)$ and $Ap_2(R-r)$ are such that $p_1(0) > Ap_2(R)$ and $p_1(R) < Ap_2(0)$ (from the corollary statement), there is a unique quantity $r^*$ such that $p_{1}(r^*) = A p_{2}(R - r^*)$. Now, substituting in the equality yields
\begin{align*}
 p_{1}(r^*) = A p_{2}(R - r^*) \iff & \exp(-r^*) =A \exp(r^* - R)\\
\stackrel{(a)}\iff & -r^* = \log A - R + r^* \\
\iff & r^*=\frac{R-\log A}{2}.    
\end{align*}
Note that (a) holds by taking logarithms for both sides.
\end{proof}

From the above corollary, it is clear that under the above assumptions, when $A=1$ the best investment strategy is $r^*=\frac{R}{2}$ (i.e., the defender half of the budget on each site). For $A<1$, the $\log A$ term is negative, hence $r^* > \frac{R}{2}$ (i.e., the defender allocates more resources on site 1). Finally, for $A>1$, we have $r^* < \frac{R}{2}$ (i.e., the defender allocates more resources on site 2 since it has higher loss valuation).

\begin{figure}
\begin{center}
\includegraphics[width=0.75\linewidth]{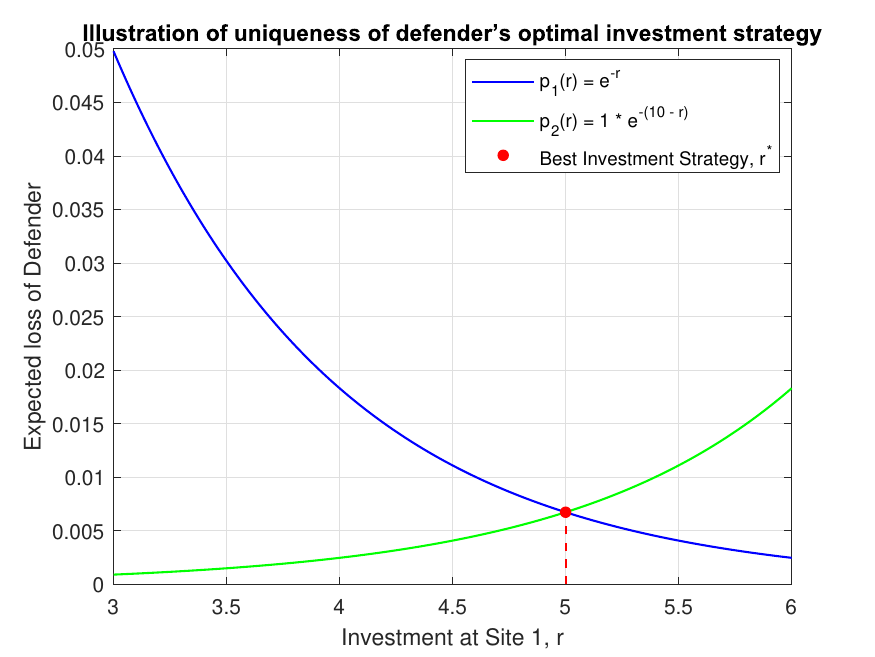}
\vspace{-1mm}
  \caption{Illustration of the defender's optimal  resource allocation that minimizes the maximum expected loss.}
  \label{fig:Sequential optimal strategy}
\end{center}
\vspace{-3mm}
\end{figure}

Fig.~\ref{fig:Sequential optimal strategy} illustrates $r^*$, i.e., the allocation that minimizes the
maximum expected loss for the defender. In this illustrative figure, we consider that the defense budget is 10 and loss at site $2$ is one, i.e., both sites incur same loss (unit loss) when compromised by the attacker. %\textcolor{red}{MA: Azim, add figure similar to Figure 2 in my CDC paper, but with $p_1(r)$ and $Ap_2(R-r)$.}\textcolor{green}{Az: Figure added}

As anticipated, the optimal approach for the defender is to select $r$ to equalize the expected loss from both sites, if feasible. While the paper~\cite{powell2007allocating} investigated this game under the assumption of perfect rationality for both the defender and the attacker, they did not delve into the effects of quantal behavioral biases exhibited by either party. On the other hand, the prior work~\cite{abdallah2020effect} focused only on the effect of non-linear probability weighting of attack success probability (from prospect theory) on the defender's investments in that sequential game setup.  In this paper, we delve into the quantal behavioral bias of the defender (where the defender is assumed to make errors in choosing which pure security investment to allocate on the sites). We outline this scenario in the following section and subsequently scrutinize the consequences of this quantal behavioral decision-making.

\section{Quantal Response Equilibrium}\label{sec:QRE_Theory}

\subsection{Quantal Response Modeling}
Quantal response equilibrium (QRE) is a solution concept in game theory which provides an equilibrium concept with bounded rationality~\cite{goeree2005regular}. It takes into account bounded rationality and stochasticity in players' decision-making processes. It provides a structural, statistical model of human operators where humans consistently make errors in choosing efficient strategies as shown in behavioral economics and psychology when modeling human decision-making (e.g.,~\cite{anderson2004noisy,zhuang2014stability}). In contrast to the deterministic and perfect best-response behavior of PNE, in the QRE players ``better respond" and choose strategies that provide higher payoffs with a higher probability.

\textbf{Incorporating QRE in Our Sequential Game:}
 In sequential security games, incorporating QRE allows for a more realistic representation of human decision-making, acknowledging that players may not always select the optimal strategies but instead exhibit a degree of randomness or error in their choices. In our setup we use logit QRE where defender's and attacker's strategies are chosen probabilistically, with the likelihood of selecting a particular strategy influenced by its expected benefit (i.e., expected loss for defender and expected gain for the attacker) relative to other available strategies and a noise parameter $\lambda$ representing the level of noise or randomness in decision-making.  This reflects the inherent unpredictability of real-world cybersecurity scenarios\cite{mckelvey1995quantal}.

We incorporate such a QRE concept into our game setup, as detailed next.

\textbf{Defender's Quantal Response:} In our sequential defender-attacker security game, the defender's security investment profiles would be chosen according to the following probability distributions (given by logit function)
\begin{equation}\label{eq:logit_response_func_defender}
\displaystyle \sigma_d (\mathbf{r}) = \frac{e^{-\lambda_d L(\mathbf{r})}}{\sum_{\mathbf{r} \in X} e^{-\lambda_d  L(\mathbf{r})}},
\end{equation}
where $\sigma_d (\mathbf{r})$ is the probability of defender choosing investment profile $\mathbf{r} \in X$, and $\lambda_d$ represents
the rationality level of the defender. Note that $L(\mathbf{r})$ is the expected loss of defender when choosing investment strategy $\mathbf{r}$ which is given by~\eqref{eq: Non_Behavioral_defender_expected_loss}.  Here, the probability $\sigma_d (\mathbf{r})$ of defender choosing security investment profile $\mathbf{r}$ increases if the defender's expected loss $L(\mathbf{r})$ decreases under that $\mathbf{r}$.

\textbf{Attacker's Quantal Response:} After observing the defender's investment, the attacker chooses the site with maximum loss to put its effort. The attacker action (which site to attack after observing defender's investments) would be chosen according to the following probability distributions %\textcolor{red}{MA: To fix this part!}\textcolor{green}{Az: added Attacker's investment profile} 
\begin{equation}\label{eq:logit_response_func_attacker}
\sigma_a (\mathbf{y}) = \frac{e^{-\lambda_a U(\mathbf{y})}}{\sum_{\mathbf{y} \in Y} e^{-\lambda_a U(\mathbf{y})}},
\end{equation}
where $\sigma_a (\mathbf{y})$ is the probability of attacker choosing attack profile $\mathbf{y} \in Y$, where  $Y := \{(1, 0), (0, 1)\}$ with the first strategy denoting attacking site $1$ and the second strategy denoting attacking site $2$, and $\lambda_a$ represents
the rationality level of the attacker.
Thus, we denote the vectors of probabilities of choosing defense  and attack strategies by $\boldsymbol\sigma_d$ and $\boldsymbol\sigma_a$, respectively. Based on such quantal responses, our search seeks to find fixed strategies to achieve quantal response equilibrium $(\boldsymbol\sigma_d^{*}, \boldsymbol\sigma_a^{*})$ as in mean field theory~\cite{anderson2004noisy}.

In our evaluation, we show the effects of the non-negative parameters $\lambda_i$, $i \in \{a,d\}$  which represent the rationality level of the player. When $\lambda_i \rightarrow 0$, the player becomes ``completely non-rational (behavioral)'' and chooses each investment profile with equal probability. As $\lambda_i \rightarrow \infty$, players become ``perfectly rational'' and the game approaches a Pure-strategy Nash equilibrium (PNE).

\subsection{Existence of a QRE}
We first establish the existence of a QRE for the class of sequential attacker-defender game defined in Section~\ref{sec:QRE_Theory}. Recall that the vector $\boldsymbol\sigma_d$ represents the likelihoods of selecting each security investment profile by the defender. These likelihoods are determined by the defender's quantal response function~\eqref{eq:logit_response_func_defender}, which describes how sensitive defender is to the differences in expected loss between different defense investment strategies. Similarly,  the vector $\boldsymbol\sigma_a$ represents the likelihoods of selecting each attack strategy by the attacker. These likelihoods are determined by the attacker's quantal response function~\eqref{eq:logit_response_func_attacker}. This motivates showing the existence of QRE in our setup in Section~\ref{sec:QRE_Theory}.

\begin{proposition}\label{prop:existence_pne}
Suppose the quantal response function for the defender is given by~\eqref{eq:logit_response_func_defender} and the  quantal response function for the attacker is given by~\eqref{eq:logit_response_func_attacker}. Suppose that the probability of successful attack $p_i(\cdot)$ has $p'_{i} < 0$ and $p''_{i} > 0$ for every site $i$. Then, our sequential security game possesses a QRE when $\lambda_d \in [0,\infty)$ and $\lambda_a \in [0,\infty)$.
\end{proposition}

\begin{proof}
Since $\lambda_d \in [0,\infty)$ and $\lambda_a \in [0,\infty)$ and   $p'_{i} < 0$ and $p''_{i} > 0$ for each site $i$, thus the logit quantal response functions in~\eqref{eq:logit_response_func_defender} and~\eqref{eq:logit_response_func_attacker} are interior, continuous, monotonic, and responsive~\cite{goeree2005regular}. Therefore, 
they are regular quantal response functions~\cite{goeree2005regular}. As a result, the sequential security game considered in our work possesses a quantal response equilibrium (QRE) \cite{mckelvey1995quantal}.
\end{proof}

Having defined our setup and proved the existence of QRE in our defender-attacker sequential games, we next show the main properties that arise under QRE.

\vspace{-2mm}

\section{Main Properties of QRE}
\vspace{-1mm}
In this section, we focus only on the properties of quantal behavioral decision-making by the defender in order to
better understand its impact on the game, and leave the consideration of a behavioral attacker for future work.
\vspace{-2mm}
\subsection{Effect of Behavioral Level $\lambda$}\label{Sub:Effect_Lambda}

We start by analyzing the effect of behavioral parameter ($\lambda_d$ for defender). In particular, we show the effect of $\lambda_d$ on the best defense strategy $\mathbf{r^*}$ for the defender.

\begin{proposition}\label{prop:effect_lambda_defender}
Suppose that the expected loss of the defender $L(\mathbf{r})$ is given by~\eqref{eq: Non_Behavioral_defender_expected_loss} and that the quantal response function for the defender is given by~\eqref{eq:logit_response_func_defender}. Let the defender's best strategy be given by $\mathbf{r}^*$. Then, the probability of choosing the best investment strategy $\mathbf{r}^*$ (denoted by $\sigma({\mathbf{r}^*})$) is non-decreasing in $\lambda_d$, 
\iffalse
the behavioral level of the defender $we have$ 
$$
\displaystyle \frac{\partial\displaystyle \sigma({\mathbf{r}^*})}{\partial \lambda_d} \geq 0,
$$
\fi
where $\lambda_d$ is the quantal behavioral level of the defender.
\end{proposition}

\begin{proof}
To find the relationship between the probability of choosing defender's best investment profile $\mathbf{r}^*$, given by $\sigma({\mathbf{r}^*})$, and the behavioral level of the defender $\lambda_d$, we need to find the partial derivative of $\sigma({\mathbf{r}^*})$ with respect to $\lambda_d$.
\vspace{-1mm}
\begin{align*}
\frac{\partial\displaystyle \sigma({\mathbf{r}^*})}{\partial \lambda_d} &=\frac{\partial }{\partial \lambda}\frac{e^{-\lambda_d L(\mathbf{r}^*)}}{\sum_{\mathbf{r} \in X} e^{-\lambda_d L(\mathbf{r})}} \\
&= \frac
{A - B}
{\left(\sum_{\mathbf{r} \in X} e^{-\lambda_d L(\mathbf{r})}\right)^2}, 
\end{align*}
where $$A = \left(\sum_{\mathbf{r} \in X} e^{-\lambda_d  L(\mathbf{r})}\right) \times \frac{\partial}{\partial \lambda}e^{-\lambda_d L(\mathbf{r}^*)},$$ and $$B  =   e^{-\lambda_d L(\mathbf{r}^*)} \times \frac{\partial }{\partial \lambda}(\sum_{\mathbf{r} \in X} e^{-\lambda_d  L(\mathbf{r}^*)}).$$

Thus, doing the differentiation, we have $A - B$ given as follows 
\vspace{-2mm}
\begin{align*}
A - B &= -L(\mathbf{r^*}) \times \left(\sum_{\mathbf{r} \in X} e^{-\lambda_d  L(\mathbf{r})}\right) \times e^{-\lambda_d L(\mathbf{r}^*)} \\
&- e^{-\lambda_d L(\mathbf{r}^*)}\times \left(\sum_{\mathbf{r} \in X} (-L(\mathbf{r}))\cdot e^{-\lambda_d  L(\mathbf{r})}\right) \\ 
&= e^{-\lambda_d L(\mathbf{r}^*)} \times \sum_{\mathbf{r} \in X} (L(\mathbf{r})-L(\mathbf{r}^*)). 
\end{align*}

Therefore, the derivative $\frac{\partial\displaystyle \sigma({\mathbf{r}^*})}{\partial\lambda_d}$ would be given by 
\vspace{-1mm}
\begin{equation}\label{eq:partial_derivative_lambda}
\frac{\partial \sigma(\mathbf{r^*})}{\partial \lambda_d} = \frac{e^{-\lambda_d L(\mathbf{r}^*)} \times \sum_{\mathbf{r} \in X} (L(\mathbf{r})-L(\mathbf{r}^*))}{\left(\sum_{\mathbf{r} \in X} e^{-\lambda_d L(\mathbf{r})}\right)^2}.
\end{equation}

Note that the exponential parts of the derivative in  \eqref{eq:partial_derivative_lambda} is always positive, hence the sign of the derivative in~\eqref{eq:partial_derivative_lambda} depends on the term $L(\mathbf{r})-L(\mathbf{r}^*)$. Since $\mathbf{r}^*$ is the defender's best strategy, then the expected loss of the defender when choosing strategy $\mathbf{r}^*$, given by $L(\mathbf{r}^*)$, is always less then or equal that under any other investment strategy of the defender. In other words, we have $L(\mathbf{r}) \geq L(\mathbf{r}^*) \forall \mathbf{r} \in X$. Thus, the numerator of~\eqref{eq:partial_derivative_lambda} is always non-negative. Since the denominator is always positive (summation of exponential terms), we have $\frac{\partial\displaystyle \sigma(\mathbf{r}^*)}{\partial \lambda_d}\geq 0$. This concludes the proof.
\end{proof}

In words, Proposition~\ref{prop:effect_lambda_defender} shows that the defender's QRE probability of choosing the best investment strategy $\mathbf{r^*}$ is non-decreasing in the  quantal behavioral level of the defender $\lambda_d$. This shows that the defender would choose the best investment strategy with higher probability when her quantal behavioral level decreases (i.e., $\lambda_d$ increases).

\subsection{Effect of Loss on QRE\label{Sub:Effect_Loss}}

We now characterize the impact of the sites' losses  on the defender's investments in each of the two sites. In particular, we show the effect of loss parameter $A$ on choosing the best strategy in the QRE by the defender. In order to keep the exposition clear, we will make the following assumption on the probabilities and site loss values for the defender (this assumption rules out the two corner cases where a defender invests entirely in only one of the two sites under her control).  This assumption was also considered in \cite{powell2007allocating}.

\begin{assumption}\label{ass: equalizing_rational}
The quantities $p_1(r)$ and $Ap_2(R-r)$ are such that $p_1(0) > Ap_2(R)$ and $p_1(R) < Ap_2(0)$, i.e., there is a unique quantity $r^*$ such that $p_{1}(r^*) = A p_{2}(R - r^*)$.
\end{assumption}

%\begin{assumption}\label{ass: equalizing_rational}

\begin{proposition}\label{prop:losses}
  Under Assumption 1, suppose that the defender's best strategy is given by $\mathbf{r}^* = [r^*, R- r^*]$. Let $p_1(r) = e^{-r}$ and $p_{2}(r)=e^{-(R-r)}$, respectively. Suppose that any other feasible defense strategy be given by $\mathbf{\hat{r}} \in X - \{\mathbf{r^*}\}$. Thus, we have the following cases.
\begin{enumerate}
%    \item If $r^*$ satisfy $p_{1}(r^*) = A p_{2}(R - r^*)$, and
        %\begin{itemize}
        \item If $ p_1(\hat{r}) > A p_2(R-\hat{r})  \forall \mathbf{\hat{r}} \in X - \{\mathbf{r}^*\}$. Suppose that there are two possible values for the loss $A$ at site $2$, given by $A_1$ and $A_2$ where $A_1 < A_2$, then we have $\sigma(\mathbf{r^*})|_{A_1} > \sigma(\mathbf{r^*})|_{A_2}$. 

    \item If $p_1(\hat{r}) < A p_2(R-\hat{r}) \forall \mathbf{\hat{r}} \in X - \{\mathbf{r}^*\}$, then we have $\frac{\partial\displaystyle \sigma({\mathbf{r}^*})}{\partial A}>0$.

%\item If $ p_1(\hat{r}) > A p_2(R-\hat{r})  \forall \hat{r} \in X$. Suppose that we have two possible values for the loss $A$ at site $2$, given by $A_1$ and $A_2$ where $A_1 < A_2 < 1$, then $\sigma(\mathbf{r^*})|_{A_1} < \sigma(\mathbf{r^*})|_{A_2}$. 
        
%        \item If $w(p_1(r)) > A w(p_2(R-r)) \forall r\in [0,R]$, then $\hat{r} \geq r^*$.
        %\item In both of the cases i(a) and i(b), if $A = 1$, then $\hat{r} = r^*$.
        %\end{itemize}
\end{enumerate}

%iii) For some $\hat{r} \in X$ has $p_1(\hat{r}) > A p_2(R-\hat{r})$ and for some $\bar{r} \in X$ has $p_1(\bar{r}) \leq A p_2(R-\bar{r}) $, where $\hat{r}$ and $\bar{r}$ all the investment strategies expect the QRE one. 
\end{proposition}
\vspace{3mm}

\begin{proof}
 We analyze the effect of loss $A$, by calculating the difference between $\sigma(\mathbf{r^*})|_{A_1}$ and $\sigma(\mathbf{r^*})|_{A_2}$ for the all different cases of the proposition. We begin with case (i): \begin{align*}
 \displaystyle \sigma({\mathbf{r}^*})|_{A_1} 
 &= \frac{e^{-\lambda_d L(\mathbf{r}^*)}}{\sum_{\mathbf{r} \in X} e^{-\lambda_d L(\mathbf{r})}} \\
&\stackrel{(a)}= \frac{e^{-\lambda_d A_1 p_2(R-r^*)}}{e^{-\lambda_d A_1 p_2(R-r^*)}+\sum\limits_{\mathbf{\hat{r}} \in X -\{\mathbf{r^*}\}}
 e^{-\lambda_d p_1(\hat{r})}}\\
&\stackrel{(b)}= \frac{e^{-\lambda_d A_1 e^{-\left(\frac{-\log(A_1)+R}{2}\right)}}}{e^{-\lambda_d A_1 e^{-\left(\frac{-\log(A_1)+R}{2}\right)}}+\sum\limits_{\mathbf{\hat{r}} \in X -\{\mathbf{r^*}\}}
 e^{-\lambda_d p_1(\hat{r})}}.
\end{align*}
Note that (a) holds since $p_{1}(r^*) = A p_{2}(R - r^*)$ and
       $ p_1(\hat{r}) > A p_2(R-\hat{r})  \forall \mathbf{\hat{r}} \in X - \{r^*\}$, and (b) holds from $r^*$ given by Corollary 1 (since $p_{1}(r^*) = A p_{2}(R - r^*)$ and from substituting $p_2(r)$ from the  proposition statement). 
       
       Similarly, $\sigma({\mathbf{r}^*})|_{A_2}$ would be 
\begin{align*}
 \displaystyle \sigma({\mathbf{r}^*})|_{A_2} 
 &= \frac{e^{-\lambda_d A_2 e^{-\left(\frac{-\log(A_2)+R}{2}\right)}}}{e^{-\lambda_d A_2 e^{-\left(\frac{-\log(A_2)+R}{2}\right)}}+\sum\limits_{\mathbf{\hat{r}} \in X -\{\mathbf{r^*}\}}
 e^{-\lambda_d p_1(\hat{r})}}.
\end{align*}

\iffalse
$$
=\frac{\left(e^{-\lambda_d A p_2(R-r^*)}+\sum_{\mathbf\hat{r} \in X} e^{-\lambda_d p_1(\hat{r})}\right)\frac{\partial}{\partial A} e^{-\lambda_d A p_2(R-r^*)}- e^{-\lambda_d A p_2(R-r^*)} \frac{\partial}{\partial A}\left(e^{-\lambda_d A p_2(R-r^*)}+\sum_{\mathbf\hat{r} \in X} e^{-\lambda_d p_1(\hat{r})}\right)}
{\left(e^{-\lambda_d A p_2(R-r^*)}+\sum_{\mathbf\hat{r} \in X} e^{-\lambda\ p_1(\hat{r})}\right)^2}
$$
\fi

%\textcolor{green}{Az: The result will be different(zero) if we use $p_{1}(r^*)$ instead of $ A p_{2}(R - r^*)$, as  $p_{1}(r^*) = A p_{2}(R - r^*)$}

Now, subtracting $\displaystyle \sigma({\mathbf{r}^*})|_{A_1}$ and $\displaystyle \sigma({\mathbf{r}^*})|_{A_2}$ would yield

\begin{align*}
 \displaystyle \sigma({\mathbf{r}^*})|_{A_1} - \sigma({\mathbf{r}^*})|_{A_2} 
 &= \frac{B - C}{D}, 
\end{align*}
where 
\begin{align*}
B - C &= \sum_{\mathbf{\hat{r}} \in X -\{\mathbf{r^*}\}} e^{-\lambda_d p_1(\hat{r})} \times \left(e^{-\lambda_d A_1 e^{-\left(\frac{-\log(A_1)+R}{2}\right)}} \right. \\
&-\left. e^{-\lambda_d A_2 e^{-\left(\frac{-\log(A_2)+R}{2}\right)}}\right), \ \text{and}\\
D &= \left(e^{-\lambda_d A_1 e^{-\left(\frac{-\log(A_1)+R}{2}\right)}}+\sum_{\mathbf{\hat{r}} \in X -\{\mathbf{r^*}\}} e^{-\lambda_d p_1(\hat{r})}\right)\\
&\times \left(e^{-\lambda_d A_2 e^{-\left(\frac{-\log(A_2)+R}{2}\right)}}+\sum_{\mathbf{\hat{r}} \in X -\{\mathbf{r^*}\}} e^{-\lambda_d p_1(\hat{r})}\right).
\end{align*}
Note that $D$ is always positive (since it is multiplication of two terms, with each term being composed of summation of several exponential terms). Now, we check the term $B - C$. %as follows.
\begin{align*}
 A_1 < A_2 
 \iff & \log(A_1) < \log(A_2)\\ 
\iff & -\log(A_1) > -\log(A_2)\\ 
\iff & R-\log(A_1) > R-\log(A_2)\\
\iff & -\left(\frac{R-\log(A_1)}{2}\right) < -\left(\frac{R-\log(A_2)}{2}\right)\\
\iff & A_1 e^{-\left(\frac{-\log(A_1)+R}{2}\right)} < A_2 e^{-\left(\frac{-\log(A_2)+R}{2}\right)} \\
\iff & e^{-\lambda_d A_1 e^{-\left(\frac{-\log(A_1)+R}{2}\right)}} \\ &> e^{-\lambda_d A_2 e^{-\left(\frac{-\log(A_2)+R}{2}\right)}}.
\end{align*} Thus, we have $B - C > 0$. Thus, $\displaystyle \sigma({\mathbf{r}^*})|_{A_1} - \sigma({\mathbf{r}^*})|_{A_2} > 0$, which concludes the case (i).

\iffalse
For the case i(b), we again analyze the term $B -C$ as follows. Following same steps as the case i(a), we have
\begin{align*}
  A_1 < A_2  < 1 \\
\iff & e^{-\lambda_d A_1 e^{-\left(\frac{-\log(A_1)+R}{2}\right)}} > e^{-\lambda_d A_1 e^{-\left(\frac{-\log(A_1)+R}{2}\right)}}.
\end{align*}
\fi

%\textcolor{red}{MA: Here, we stopped in proofing first part of Proposition 4, rest is not related}

 Case (ii):  We do this case by differentiating $\sigma({\mathbf{r}^*})$ with respect to $A$ as follows.
\begin{align*}
 \frac{\partial\displaystyle \sigma({\mathbf{r}^*})}{\partial A} &=\frac{\partial }{\partial A}\frac{e^{-\lambda_d L(\mathbf{r}^*)}}{\sum_{\mathbf{r} \in X} e^{-\lambda_d L(\mathbf{r})}} \\
&\stackrel{(c)}= \frac{\partial }{\partial A}\frac{e^{-\lambda_d A p_2(R-r^*)}}{e^{-\lambda_d A_1 p_2(R-r^*)}+\sum\limits_{\mathbf{\hat{r}} \in X -\{\mathbf{r^*}\}}
 e^{-\lambda_d A p_2(R-\hat{r})}}\\
&= \frac{E - F}{G},
\end{align*}
Note that (c) holds since $p_{1}(r^*) = A p_{2}(R - r^*)$ and $ p_1(\hat{r}) < A p_2(R-\hat{r})  \forall \mathbf{\hat{r}} \in X - \{\mathbf{r^*}\}$. Now, we give the expressions for $E$, $F$, and $G$, which are given as follows. 
\begin{align*}
E &= \bigg(e^{-\lambda_d A p_2(R-r^*)}+\sum_{\mathbf{\hat{r}}\in X-\{\mathbf{r^*}\}} e^{-\lambda_d A p_2(R-\hat{r})}\bigg)\\ &\times \frac{\partial}{\partial A} e^{-\lambda_d A p_2(R-r^*)}, \\
F &= e^{-\lambda_d A p_2(R-r^*)} \frac{\partial}{\partial A}\bigg(e^{-\lambda_d A p_2(R-r^*)} \\
& + \sum_{\mathbf{\hat{r}}\in X-\{\mathbf{r^*}\}} e^{-\lambda_d A p_2(R-\hat{r})}\bigg),
\end{align*}
and $G = {\left(e^{-\lambda_d A p_2(R-r^*)}+\sum_{\mathbf{\hat{r}}\in X-\{\mathbf{r^*}\}} e^{-\lambda_d A p_2(R-\hat{r})}\right)^2}$.
Since the denominator $G$ is always positive, we explore the numerator part $E - F$ to check its sign as follows. 
\begin{align*}
E - F &=\left(\sum_{\mathbf{\hat{r}}\in X-\{\mathbf{r^*}\}} e^{-\lambda_d A p_2(R-\hat{r})}\right) \frac{\partial}{\partial A} e^{-\lambda_d A p_2(R-r^*)} \\
&- e^{-\lambda_d A p_2(R-r^*)} \frac{\partial}{\partial A} \left(\sum_{{\mathbf{\hat{r}}\in X-\{\mathbf{r^*}\}}} e^{-\lambda_d A p_2(R-\hat{r})}\right).
\end{align*}
Note that  
\begin{align*}
\frac{\partial}{\partial A} e^{-\lambda_d A p_2(R-r^*)} &=  \left(p_2(R-r^*)+A\frac{\partial p_2(R-r^*)}{\partial A}\right)\\
&\times e^{-\lambda_d A p_2(R-r^*)} \times (-\lambda_d)\\
&\stackrel{(d)}= -\lambda_d \frac{p_2(R - r^*)}{2} e^{-\lambda_d A p_2(R-r^*)}.
\end{align*}
Note that (d) holds since the differentiating $\frac{\partial p_2(R-r^*)}{\partial A}$ would be given by  $\frac{\partial p_2(R-r^*)}{\partial A} 
    = -\frac{p_2(R-r^*)}{2A}$ (using the equality $p_1(r^*)=A \ p_2(R-r^*)$ in Assumption 1 and $r^*$ from Corollary 1). Moreover, we have 
$$
\hspace{-30mm}    \frac{\partial}{\partial A} \left(\sum_{{\mathbf{\hat{r}}\in X-\{\mathbf{r^*}\}}} e^{-\lambda_d A p_2(R-\hat{r})}\right) 
 $$
 $$
    = - \sum_{{\mathbf{\hat{r}}\in X-\{\mathbf{r^*}\}}} \lambda_d p_2(R - \hat{r}) e^{-\lambda_d A p_2(R-\hat{r})}.
$$
Substituting with the above two terms in $E - F$, we have
\begin{align}\label{eq:partial_derivative_Loss_case 2}
E - F &= \lambda_d e^{-\lambda_d A p_2(R-r^*)} \nonumber \\
&\hspace{-10mm}\times     \sum_{\mathbf{\hat{r}} \in X -\{\mathbf{r^*}\}} e^{-\lambda_d A p_2(R-\hat{r})}\left(p_2(R-\hat{r})-\frac{p_2(R-r^*)}{2}\right).
\end{align}
Note that the exponential terms and $\lambda_d$ in equation~\eqref{eq:partial_derivative_Loss_case 2} are always positive. Hence, the sign of the derivative $\frac{\partial\displaystyle \sigma({\mathbf{r}^*})}{\partial A}$ depends on the term $p_2(R-\hat{r})-\frac{p_2(R-r^*)}{2}$. Since $r^*$ is the defender's unique best strategy in the QRE (by Assumption 1), the probability of successful attack for QRE strategy $r^*$ is always less than the probability of successful attack for any other strategy $\hat{r}$ for the defender. Hence, we have
\begin{align*}
 p_2(R-\hat{r}) > p_2(R-r^*)
 \iff &p_2(R-\hat{r})>\frac{p_2(R-r^*)}{2}. 
\end{align*}
Thus, we have $E - F > 0$, which yields that $\frac{\partial\displaystyle \sigma({\mathbf{r}^*})}{\partial A} > 0$.

\end{proof}

Proposition~\ref{prop:losses} shows that the defender's choice of the  best strategy at the quantal response equilibrium  increases with the increase of the loss of the second site $A$ when $p_1(\hat{r}) < A p_2(R-\hat{r})$, where $\hat{r}$ are all the strategies except the optimal strategy 
 (i.e., when other strategies under-invest defense budget on site 2). On the other hand, it decreases with the loss of the second site $A$ if $p_1(\hat{r}) > A p_2(R-\hat{r})$ (i.e., when other strategies over-invest budget on site 2).

\section{Measuring Inefficiency of QRE: The Price of Quantal Response}\label{sec:PoBM}

The notion of Price of Anarchy (PoA) is commonly utilized to assess the inefficiencies of a Nash equilibrium when compared to the socially optimal outcome\cite{roughgarden2003price}. 
%More precisely, the Price of Anarchy is the measure of the highest total system cost at a Pure Nash Equilibrium (PNE) relative to the total system cost at the socially optimal state. 
In our context, we aim to establish a metric that accounts for inefficiencies in the equilibrium resulting from the individual strategic behaviors of defender and its behavioral decision-making characterized by quantal errors. Therefore, we introduce the concept of Price of Quantal Anarchy (PoQA), which quantifies the ratio of the total expected cost of the defender when considering defender's behavioral choices at the quantal response equilibrium in comparison to the minimum possible loss of the non-behavioral defender (that always chooses optimal defense strategy).

Specifically, we define $L(\mathbf{r}^*) \triangleq \underset{\mathbf{r} \in X}{\text{Minimum}} \enspace L(\mathbf{r})$, where $L(\mathbf{r}^*)$ (where $L(\cdot)$ is defined in \eqref{eq: Non_Behavioral_defender_expected_loss}) is the minimum loss suffered by defender under optimal investment vector $\mathbf{r^*}$. Let $X^{\mathtt{QRE}}$ denotes the set of all possible investments at QRE for the defender. We now define the Price of Quantal Anarchy as
\begin{equation}\label{eq:PoQA_QRE_Defender}
     PoQA = \frac{\sum_{\mathbf{r} \in X^\mathtt{QRE}} \sigma(\mathbf{r}) L(\mathbf{r})}{L(\mathbf{r}^{*})},
\end{equation}
where each expected loss $L(\mathbf{r})$ for defender in the numerator under investment profile $\mathbf{r} \in X^{QRE}$ is weighted with the probability $\sigma(\mathbf{r})$ of choosing that investment profile, and $\mathbf{r^*} = [r^*, R - r^*]$ with $r^* = \displaystyle \min_{r\in [0,R]} \max\{p_{1}(r), A p_{2}(R - r)\}$ is the optimal investment profile that yields the minimum expected loss of the rational defender.
In our evaluation, we also refer to the PoQA as the ``inefficiency'' of the QRE.

\subsection{Bounds on the PoQA}
We now establish an upper bound on the PoQA. In particular, we show that the PoQA is bounded if the total budget is bounded (regardless of the defenders' behavioral level $\lambda_d$). We show such bound with the following proposition.

\begin{proposition}\label{prop:POQA_upper_bound}
Let the budget available to the defender be $R$, and let the probability of successful attack on sites $1$ and $2$ be given by $p_{1}(r)=e^{-r}$ and $p_{2}(r)=e^{-(R-r)}$, respectively. Let $X^{\mathtt{QRE}}$ denotes the set of all possible investments at QRE for the defender. Then, for any $A$ and any behavioral level $\lambda_d \in [0,\infty)$,  $PoQA\leq \max\{A, \frac{1}{A}\} |X^{\mathtt{QRE}}| \exp(R)$. 

\end{proposition}
\begin{proof}
Starting with the PoQA equation in~\eqref{eq:PoQA_QRE_Defender}, we have
\begin{align*}
PoQA &= \frac{\sum_{\mathbf{r} \in X^\mathtt{QRE}} \sigma(\mathbf{r}) L(\mathbf{r})}{L(\mathbf{r}^{*})},\\
&\stackrel{(a)} \leq \frac{\sum_{\mathbf{r} \in X^\mathtt{QRE}} L(\mathbf{r})}{L(\mathbf{r}^{*})}\\
&\stackrel{(b)} \leq \frac{ |X^\mathtt{QRE}| \times L(\mathbf{\hat{r}})}{{L(\mathbf{r}^{*})}},
\end{align*}
where $L(\mathbf{\hat{r}}) \triangleq \underset{\mathbf{r} \in X^\mathtt{QRE}}{\text{max}} \enspace L(\mathbf{r})$. Note that (a) holds since the probability of choosing investment profile $\mathbf{r}$ is upper bounded by $1$, i.e.,  $\sigma(\mathbf{r}) \in (0,1]$, and that $(b)$ holds from the definition of $L(\mathbf{\hat{r}})$, which is the worst case defender's loss.

Now, substituting from \eqref{eq: Non_Behavioral_defender_expected_loss} into the above bound, we have
\begin{align*}
PoQA &\leq \frac{ |X^\mathtt{QRE}| \times \text{max} \left\{p_{1}(\hat{r}), A p_{2}(R - \hat{r})\right\}}{\text{max} \left\{p_{1}(r^*), A p_{2}(R - r^*)\right\}}. 
\end{align*}

Note that  $e^{-R} \leq p_1(r) \leq 1$ and $e^{-R} \leq p_2(r) \leq 1$ from the proposition statement. We thus consider the following four possible sub-cases as follows.

i) If $p_1(\hat{r}) > A p_2(R-\hat{r})$ and $p_1(r^*) > A p_2(R-r^*)$, then we have
\begin{align*}
PoQA &\leq \frac{|X^\mathtt{QRE}| \times p_1(\hat{r})}{p_1(r^*)} \\ &\stackrel{(c)}{=} \frac{|X^\mathtt{QRE}| \times \exp(-\hat{r})}{\exp(-r^*)} \stackrel{(d)}{\leq} |X^\mathtt{QRE}| \exp(R).
\end{align*}

Note that (c) holds from the proposition statement and (d) holds since the numerator is upper bounded by $|X^\mathtt{QRE}|$  and the denominator is lower bounded by $e^{-R}$.

ii) If $p_1(\hat{r}) \leq A p_2(R-\hat{r})$ and $p_1(r^*) \leq A p_2(R-r^*)$, then we have

\begin{align*}
PoQA &{\leq} \frac{ |X^\mathtt{QRE}| \times Ap_2(R - \hat{r})}{Ap_2(R - r^*)} \\ &= \frac{|X^\mathtt{QRE}| A \exp(-R + \hat{r})}{A \exp(-R + r^*)}
{\leq} |X^\mathtt{QRE}| \exp(R).
\end{align*}

iii) If $p_1(\hat{r}) > A p_2(R-\hat{r})$ and $p_1(r^*) \leq A p_2(R-r^*)$, then we have  
\begin{align*}
PoQA &{\leq} \frac{|X^\mathtt{QRE}| \times p_1(\hat{r})}{Ap_2(R - r^*)} \\ &= \frac{|X^\mathtt{QRE}| \exp(-\hat{r})}{A\exp(-R + r^*)} \stackrel{(e)}{\leq} \frac{|X^\mathtt{QRE}| \exp(R)}{A}.
\end{align*}

Note that (e) holds since the numerator is upper bounded by $|X^\mathtt{QRE}|$  and the denominator is lower bounded by $A e^{-R}$.

iv) If $p_1(\hat{r}) \leq A p_2(R-\hat{r})$ and $p_1(r^*) > A p_2(R-r^*)$, then we have 
\begin{align*}
PoQA &{\leq} \frac{A \times |X^\mathtt{QRE}| \times p_2(R - \hat{r})}{p_1(r^*)} \\&= \frac{A|X^\mathtt{QRE}|\exp(-R + \hat{r})}{\exp(-r^*)} \stackrel{(f)}{\leq} A|X^\mathtt{QRE}|\exp(R).
\end{align*}

Similarly, (f) holds since the numerator is upper bounded by $A |X^\mathtt{QRE}|$
and the denominator is lower bounded by $e^{-R}$.

In all of the possible scenarios (as shown above), $PoQA \leq \max\{A, \frac{1}{A}\} |X^{\mathtt{QRE}}| \exp(R)$, which concludes the proof.
\end{proof}

Now, we illustrate our characterizations of the impacts of quantal decision-making via numerical simulations.

\section{Numerical Simulations}

In this section, we show our main numerical simulation results for our defender-attacker sequential game described in Section~\ref{sec:background} and the QRE setup described in Section~\ref{sec:QRE_Theory}. 
\vspace{-1mm}
\subsection{Experimental Setup}
For our simulations, we consider three different defense strategy spaces (shown in Table~\ref{tab:Invst_Strat}) which are chosen such that $\mathbf{r_3}$ remains the defender's best strategy for all spaces. These defense strategy spaces are chosen such that it can help in illustrating our theoretical results of our quantal response analysis.\footnote{Note that all our theoretical results consider the set $X$ of security investment strategies that have all possible feasible allocations.} We consider the sequential game with a behavioral defender and a non-behavioral attacker. We let the probability of successful attack on sites $1$ and $2$ be given by $p_{1}(r)=e^{-r}$ and $p_{2}(r)=e^{-(R-r)}$, respectively, where $r$ is the investment on site $1$. For the defender, the first site has a unit loss while second  site has a loss of $A$ when compromised. We let the total defense budget for defending the two sites be $R = 10$. Thus, $\mathbf{r_3}$ is the optimal strategy for each of the spaces hence $p_{1}(r_3) = A p_{2}(R - r_3)$. We consider spaces A and B such that, except for the optimal strategy $\mathbf{r_3}$, the condition $p_{1}(r_i)> A p_{2}(R - r_i)$ holds in space A, while in space B, $p_{1}(r_i) < A p_{2}(R - r_i)$. Finally, defense strategy space C includes investment strategies satisfying conditions from both spaces A and B, with one optimal strategy emerging when loss $A=1$.

\subsection{Effect of Behavioral Level (Quantal Bias Parameter)}
\begin{figure}[t!]
%\vspace{-1mm}
\centering
\includegraphics[width=0.75\linewidth]{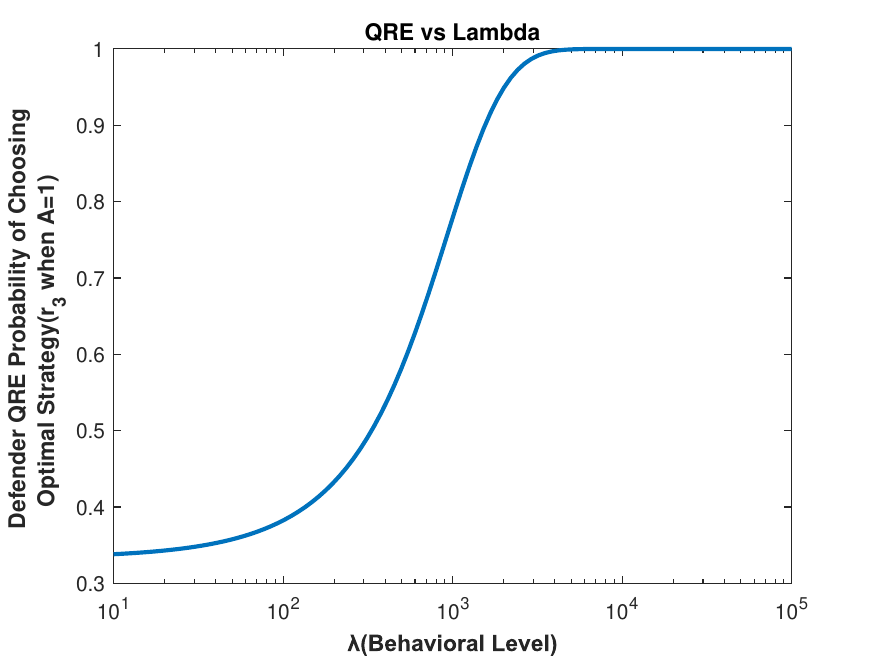}
  \caption{Effect of behavioral level $\lambda$ on the defender's QRE probability of choosing the optimal strategy. In this case, we use defense strategy space C where the loss $A=1$.} \label{fig:QRE_Sequential_Var_Lambda}
  \vspace{-2mm}
\end{figure}

Figure~\ref{fig:QRE_Sequential_Var_Lambda} shows the effect of behavioral level $\lambda_d = \lambda$ on the QRE probability of choosing optimal defense strategy $\mathbf{r_3}$ from the defense strategy space C of Table~\ref{tab:Invst_Strat}. Figure~\ref{fig:QRE_Sequential_Var_Lambda} shows that with the increase of $\lambda$ the QRE probability of the optimal defense strategy also increases which corresponds to our analysis (Proposition~\ref{prop:effect_lambda_defender}) in Section~\ref{Sub:Effect_Lambda}. This shows that a defender tends to choose better defense strategy as her quantal behavioral bias decreases (i.e., $\lambda$ increases). Also, it is evident from Figure~\ref{fig:QRE_Sequential_Var_Lambda} that  the QRE probability of choosing strategy $\mathbf{r_3}$ approaches to one for higher $\lambda$ for symmetrical losses of two sites ($A=1$), i.e., approaching the PNE.
\subsection{Effect of Loss Values of Sites}
\begin{figure}
\centering
\includegraphics[width=0.85\linewidth]{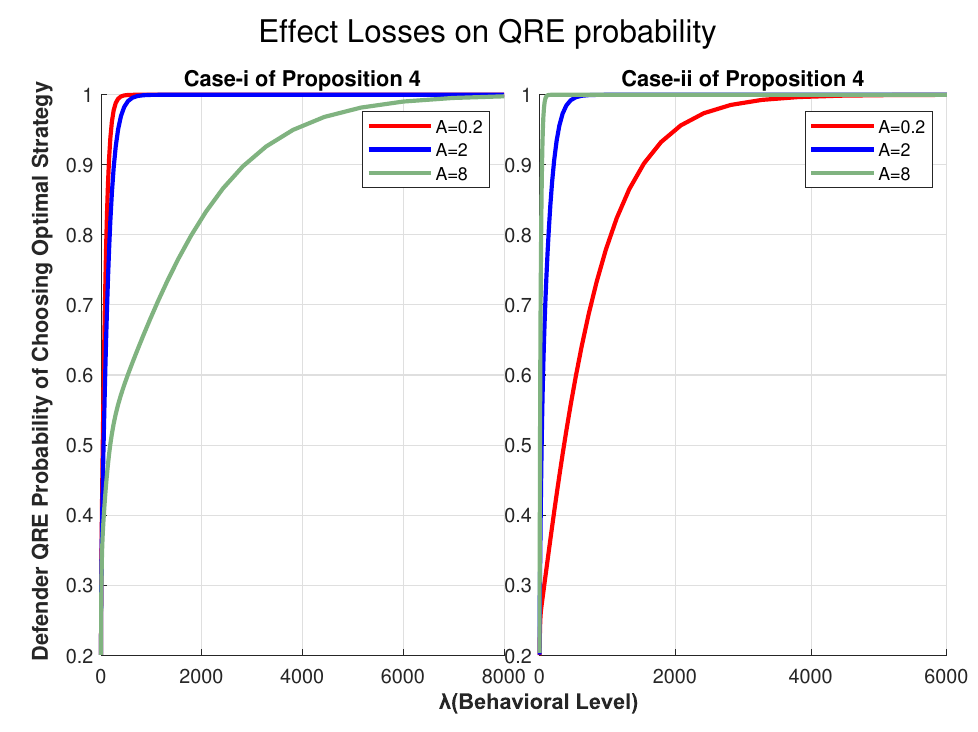}
\vspace{-1mm}
  \caption{QRE probabilities of defender's optimal strategy under different losses and strategy spaces. Left and right subplots use defense strategy spaces A and B, respectively.} 
\label{fig:QRE_Sequential_Var_Loss}
\vspace{-3mm}
\end{figure}

Figure~\ref{fig:QRE_Sequential_Var_Loss} illustrates the effect of increasing losses $\forall A>0$. We plot the left and right subplots of Figure~\ref{fig:QRE_Sequential_Var_Loss} using the defense strategy spaces A and B (from Table~\ref{tab:Invst_Strat}), respectively.  The left subplot of Figure~\ref{fig:QRE_Sequential_Var_Loss} shows that for the same behavioral level (same $\lambda$) the QRE probability of choosing the optimal strategy is decreasing when increasing the loss $A$. This result corresponds to the case-i of Proposition~\ref{prop:losses}.  The right subplot of Figure~\ref{fig:QRE_Sequential_Var_Loss} illustrates the different relationship between loss and QRE probability of choosing the optimal strategy where the QRE probability is higher for higher loss $A$. This result of right subplot of Figure~\ref{fig:QRE_Sequential_Var_Loss} is also consistent with the case-ii of Proposition~\ref{prop:losses}.

\subsection{Inefficiency of Quantal Response (PoQA)}
\begin{figure}
\centering
  \includegraphics[width=0.75\linewidth]{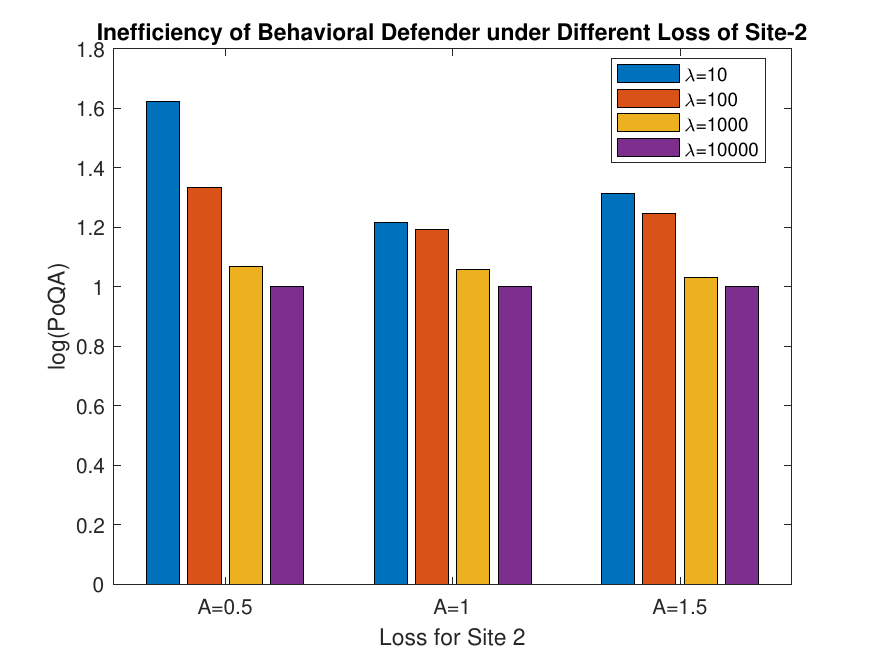}
  \caption{The inefficiency (PoQA) of behavioral defender with different behavioral levels and different losses. We show the $\log (PoQA)$ for better readability.}
  \label{fig:QRE_Sequential_PoQA}
  \vspace{-6mm}
\end{figure}

Figure~\ref{fig:QRE_Sequential_PoQA} shows the logarithm of the value of the PoQA metric (given by \eqref{eq:PoQA_QRE_Defender}) to measure the inefficiency due to behavioral defender with quantal errors, compared to the investments of a rational defender. Here, we vary $\lambda$ for the defender from 10 (highly behavioral) to 10,000 (almost rational) for different loss values of site $2$ ($A=0.5, A=1$, and $A=1.5$). Figure~\ref{fig:QRE_Sequential_PoQA} shows that the inefficiency due to behavioral decision making become worst when $A=0.5$. This is consistent with our finding in Proposition~\ref{prop:POQA_upper_bound} (recall the term $\max\{A, \frac{1}{A}\}$ in the upper bound). This term is highest for $A=0.5$ among all three illustrated loss values. This considerable increase in the PoQA indicates that the behavioral defender's investments are beneficial to the attacker, especially under loss asymmetries (i.e., one site is more valuable to the defender). This will lead eventually to higher expected loss to the defender when attacked.

\begin{table}
\vspace{2mm}
\caption{Three defense strategy spaces in our sequential game setup in Figure~\ref{fig:Sequential game setup}. Note that $A$ is the loss of site 2.} 
\label{tab:Invst_Strat}
\vspace{-2mm}
    \centering
    \small
    \setlength{\tabcolsep}{4pt} % Adjust column spacing
    \adjustbox{max width=\columnwidth}{%
    \begin{tabularx}{\linewidth}{|*{7}{>{\centering\arraybackslash}X|}} \hline 
         &  \multicolumn{2}{c|}{Space-A }&  \multicolumn{2}{c|}{Space-B }&  \multicolumn{2}{c|}{ Space-C}\\ \hline 
         Strategy&  Site 1&  Site 2&  Site 1&  Site 2& Site 1 & Site 2\\ \hline 
         $\mathbf{r_1}$&  4&  6&  7&  3&  10& 0\\ \hline 
         $\mathbf{r_2}$&  2&  8&  5.4&  4.6&  5.35& 4.65\\ \hline 
         $\mathbf{r_3}$&  $\frac{10-\log A}{2}$&  $\frac{10+\log A}{2}$&  $\frac{10-\log A}{2}$&  $\frac{10+\log A}{2}$&  5& 5\\ \hline 
         $\mathbf{r_4}$&  3.5&  6.5&  6.5&  3.5&  4.8& 5.2\\ \hline 
         $\mathbf{r_5}$&  0&  10&  10&  0&  0& 10\\ \hline
    \end{tabularx}%
    }
    \vspace{-4mm}
\end{table}

\section{Conclusion}
In this paper, we analyzed a sequential defender-attacker game dynamics through a quantal response analysis, where the defender exhibits quantal behavioral bias in investment decision. We demonstrated that as the defender's quantal behavioral bias increases, the probability of choosing the most efficient investment decreases, resulting in higher expected losses for defender and consequently higher expected gain for attacker. We showed how the sites' loss values influence the defender's choice of optimal defense investment under quantal bias. We quantified the inefficiency from quantal decision-making via introducing a metric, named the Price of Quantal Anarchy, to capture this inefficiency. We provided bounds for that metric. Our theoretical findings were confirmed through numerical simulations. Future work could extend this framework to scenarios with more sites and investigate the dynamics between non-behavioral defenders and behavioral attackers.
\vspace{-2.5mm}

\bibliographystyle{IEEEtran}
\bibliography{refs}

\begin{thebibliography}{10}
\providecommand{\url}[1]{#1}
\csname url@samestyle\endcsname
\providecommand{\newblock}{\relax}
\providecommand{\bibinfo}[2]{#2}
\providecommand{\BIBentrySTDinterwordspacing}{\spaceskip=0pt\relax}
\providecommand{\BIBentryALTinterwordstretchfactor}{4}
\providecommand{\BIBentryALTinterwordspacing}{\spaceskip=\fontdimen2\font plus
\BIBentryALTinterwordstretchfactor\fontdimen3\font minus \fontdimen4\font\relax}
\providecommand{\BIBforeignlanguage}[2]{{%
\expandafter\ifx\csname l@#1\endcsname\relax
\typeout{** WARNING: IEEEtran.bst: No hyphenation pattern has been}%
\typeout{** loaded for the language `#1'. Using the pattern for}%
\typeout{** the default language instead.}%
\else
\language=\csname l@#1\endcsname
\fi
#2}}
\providecommand{\BIBdecl}{\relax}
\BIBdecl

\bibitem{humayed2017cyber}
A.~Humayed, J.~Lin, F.~Li, and B.~Luo, ``Cyber-physical systems security -- a survey,'' \emph{IEEE Internet of Things Journal}, vol.~4, no.~6, pp. 1802--1831, 2017.

\bibitem{laszka2015survey}
A.~Laszka, M.~Felegyhazi, and L.~Buttyan, ``A survey of interdependent information security games,'' \emph{ACM Computing Surveys (CSUR)}, vol.~47, no.~2, p.~23, 2015.

\bibitem{alpcan2010network}
T.~Alpcan and T.~Ba{\c{s}}ar, \emph{Network security: A decision and game-theoretic approach}.\hskip 1em plus 0.5em minus 0.4em\relax Cambridge University Press, 2010.

\bibitem{farrow2007economics}
S.~Farrow, ``The economics of homeland security expenditures: Foundational expected cost-effectiveness approaches,'' \emph{Contemporary Economic Policy}, vol.~25, no.~1, pp. 14--26, 2007.

\bibitem{guan2017modeling}
P.~Guan, M.~He, J.~Zhuang, and S.~C. Hora, ``Modeling a multitarget attacker--defender game with budget constraints,'' \emph{Decision Analysis}, vol.~14, no.~2, pp. 87--107, 2017.

\bibitem{powell2007allocating}
R.~Powell, ``Allocating defensive resources with private information about vulnerability,'' \emph{American Political Science Review}, vol. 101, no.~4, pp. 799--809, 2007.

\bibitem{dhami2016foundations}
S.~Dhami, \emph{The foundations of behavioral economic analysis}.\hskip 1em plus 0.5em minus 0.4em\relax Oxford University Press, 2016.

\bibitem{anderson2004noisy}
S.~P. Anderson, J.~K. Goeree, and C.~A. Holt, ``Noisy directional learning and the logit equilibrium,'' \emph{The Scandinavian Journal of Economics}, vol. 106, no.~3, pp. 581--602, 2004.

\bibitem{zhuang2014stability}
Q.~Zhuang, Z.~Di, and J.~Wu, ``Stability of mixed-strategy-based iterative logit quantal response dynamics in game theory,'' \emph{PLoS One}, vol.~9, no.~8, p. e105391, 2014.

\bibitem{morgan1992analysis}
B.~J. Morgan, \emph{Analysis of quantal response data}.\hskip 1em plus 0.5em minus 0.4em\relax CRC Press, 1992.

\bibitem{anderson2007information}
R.~Anderson and T.~Moore, ``Information security economics--and beyond,'' in \emph{Annual International Cryptology Conference}.\hskip 1em plus 0.5em minus 0.4em\relax Springer, 2007, pp. 68--91.

\bibitem{redmiles2018dancing}
E.~M. Redmiles, M.~L. Mazurek, and J.~P. Dickerson, ``Dancing pigs or externalities?: Measuring the rationality of security decisions,'' in \emph{Proceedings of the 2018 ACM Conference on Economics and Computation}.\hskip 1em plus 0.5em minus 0.4em\relax ACM, 2018, pp. 215--232.

\bibitem{baddeley2011information}
M.~Baddeley, ``Information security: Lessons from behavioural economics,'' in \emph{Workshop on the Economics of Information Security}, 2011.

\bibitem{8814307}
M.~{Abdallah}, P.~{Naghizadeh}, A.~R. {Hota}, T.~{Cason}, S.~{Bagchi}, and S.~{Sundaram}, ``The impacts of behavioral probability weighting on security investments in interdependent systems,'' in \emph{2019 American Control Conference (ACC)}, July 2019, pp. 5260--5265.

\bibitem{9030279}
M.~{Abdallah}, P.~{Naghizadeh}, T.~{Cason}, S.~{Bagchi}, and S.~{Sundaram}, ``Protecting assets with heterogeneous valuations under behavioral probability weighting,'' in \emph{2019 IEEE 58th Conference on Decision and Control (CDC)}, 2019, pp. 5374--5379.

\bibitem{sanjab2017prospect}
A.~Sanjab, W.~Saad, and T.~Ba{\c{s}}ar, ``Prospect theory for enhanced cyber-physical security of drone delivery systems: A network interdiction game,'' in \emph{Communications (ICC), 2017 IEEE International Conference on}.\hskip 1em plus 0.5em minus 0.4em\relax IEEE, 2017, pp. 1--6.

\bibitem{abdallah2020effect}
M.~Abdallah, T.~Cason, S.~Bagchi, and S.~Sundaram, ``The effect of behavioral probability weighting in a sequential defender-attacker game,'' in \emph{2020 59th IEEE Conference on Decision and Control (CDC)}.\hskip 1em plus 0.5em minus 0.4em\relax IEEE, 2020, pp. 3255--3260.

\bibitem{9654432}
------, ``The effect of behavioral probability weighting in a simultaneous multi-target attacker-defender game,'' in \emph{2021 European Control Conference (ECC)}, 2021, pp. 933--938.

\bibitem{goeree2005regular}
J.~K. Goeree, C.~A. Holt, and T.~R. Palfrey, ``Regular quantal response equilibrium,'' \emph{Experimental economics}, vol.~8, pp. 347--367, 2005.

\bibitem{mckelvey1995quantal}
R.~D. McKelvey and T.~R. Palfrey, ``Quantal response equilibria for normal form games,'' \emph{Games and economic behavior}, vol.~10, no.~1, pp. 6--38, 1995.

\bibitem{roughgarden2003price}
T.~Roughgarden, ``The price of anarchy is independent of the network topology,'' \emph{Journal of Computer and System Sciences}, vol.~67, no.~2, pp. 341--364, 2003.

\end{thebibliography}

\end{document}